# THE INFLUENCE OF STELLAR FLARE ON DYNAMICAL STATE OF THE ATMOSPHERE OF EXOPLANET HD 209458b

**D. V. Bisikalo[1], A.A. Cherenkov[1], V.I. Shematovich[1], L. Fossati[2], C. Möstl[2]**

[1]*Institute of Astronomy, Russian Academy of Sciences, Moscow, Russia*

[2]*Space Research Institute, Austrian Academy of Sciences, Graz, Austria*

**Abstract.** By applying an one-dimensional aeronomic model of the upper atmosphere of the close-in giant planet HD 209458b, we study the reaction of the planetary atmosphere to an additional heating caused by the influence of a stellar flare. It is shown that the absorption of additional energy of the stellar flare in the extreme ultraviolet leads to local atmospheric heating, accompanied by formation of two shock waves, propagating in the atmosphere. We discuss possible observational manifestations of the shocks and feasibility of their detection.

INTRODUCTION

Hot-Jupiters and -Neptunes are exoplanets with masses and radii close to the parameters of Jupiter and Neptune in the Solar System, but with much higher thermospheric temperatures, up to $10^4$ K [1]. High thermospheric temperature is caused by absorption of the high-energy stellar flux, which is very strong for close-in planets. As a rule, the semi-major axes of their orbits do not exceed 0.1 a.u. It is widely agreed (see, for instance, [2]) that the atmospheric heating is caused by both stellar irradiation and interaction of the atmosphere with the stellar wind [3].

An important area of exoplanets research is the study of the influence of extreme stellar activity (flares, coronal mass ejections) on the evolution of the atmosphere of hot-Jupiters, -Neptunes, super- and exo-Earths, in particular, evaluating the mass loss from their atmospheres due to sporadic stellar phenomena (see, for instance, recent studies [4-6]). In the works [6,7], the impact of coronal mass ejections (CME) on the atmosphere of the hot-Jupiter HD 209458b was studied using a three-dimensional numerical simulation of the gas dynamics of the hot-Jupiter atmosphere during a CME passage. Characteristic changes of the flow structure of quasi-closed and closed (but significantly disturbed by the gravitational influence of the host star) gaseous envelopes of the exoplanet were derived. It was shown that a typical CME is sufficient to tear off and carry away the outer parts of the asymmetric gas envelope if it extends outside the Roche lobe.

This leads to a significant increase of the mass loss rate by the exoplanet envelope during a CME passage. However, in these studies it was assumed that the parameters of the exoplanet atmosphere do not change during extreme stellar events.

It is evident that in the case of a stellar flare, the exoplanet atmosphere reacts on the extra absorption of extreme-UV radiation and consequent heating of the atmospheric gas causes a change of the dynamic state of the atmosphere. Recently developed one-dimensional aeronomic model of the upper atmosphere of hot-Jupiters and -Neptunes [9] allows us to take into account these impacts of stellar flares on the state of the atmosphere. This model is self-consistent and takes into account both heating of the atmospheric gas due to absorption of stellar radiation in the soft X-ray and extreme ultraviolet bands by the hydrogen-dominated upper atmosphere and reactions involving suprathermal photoelectrons. The model allows to compute the height profiles of density, velocity, and temperature in giant planets atmospheres.

In this paper, using an one-dimensional aeronomic model of a giant planet with $H_2$, He-dominated atmosphere, we consider reaction of atmosphere of a hot-Jupiter HD 209458b to additional heating, caused by the influence of a stellar flare. The paper is organized as follows: in section 2 the model is described, in section 3 the results of calculations are presented. In the closing section 4, we discuss observational manifestations of the dynamic response of the hot-Jupiter atmosphere to the action of stellar flare and present our conclusions.

THE MODEL

One of the key factors determining the state of an exoplanet atmosphere is heating by the radiation of the host star. It is especially important for hot-Jupiters and -Neptunes, i.e., close-in giant planets with small (< 0.1 a.u.) orbital distances. After the discovery of the first planets of this type, it was found that the atmospheres of some of them extend beyond the limits of the Roche lobe, leading to a powerful gas-dynamic outflow of the atmospheric gas [10,11]. The heating of the hydrogen upper atmosphere is caused by the absorption of the extreme UV radiation of the host star in the 1-115 nm range.

Radiation is predominantly absorbed during ionization of atomic hydrogen and helium, as well as ionization, dissociation and dissociative ionization of molecular hydrogen. Respectively, efficiency of the heating is defined as the ratio of the total rate of local heating of atmospheric gas and the absorption rate of stellar radiation energy. This parameter is important for close-in exoplanets, subject to intense irradiation by extreme UV and soft X-ray. For instance, in the calculations of the heating of the hot-Jupiter HD 209458b [12] it has been found that, if the influence of photoelectrons is taken into account, the heating efficiency at thermospheric altitudes does not exceed 0.25.

Depending on the upper atmosphere composition and heating efficiency, the atmospheric escape regime can vary from hydrostatic to hydrodynamic. In order to study this problem, a one-dimensional self-consistent model of the hot Jupiter was developed [9]. It includes three main modules: a Monte Carlo module, a module of chemical kinetics and a gas-dynamic. In the Monte Carlo module the heating rate of the atmosphere, rates of photolytic reactions, as well as the rate of reactions caused by suprathermal photoelectrons are computed by solving the Boltzmann equation [12]. In the module of chemical kinetics, using rates of photolytic reactions obtained in the Monte-Carlo module, the system of equations of chemical kinetics is solved and concentration of atmospheric components in every cell is calculated. In the gas-dynamic module, the profiles of macroscopic parameters of the atmosphere - density, velocity, and temperature – are computed. The contribution of photoelectrons was taken into account as a function of atmospheric heating. This one-dimensional model was applied to calculate the temperature, pressure, and composition profiles of the exoplanet atmosphere at the substellar point. The gravitational potential was set equal to the three-dimensional Roche potential along the line connecting the planetary and the stellar centers.

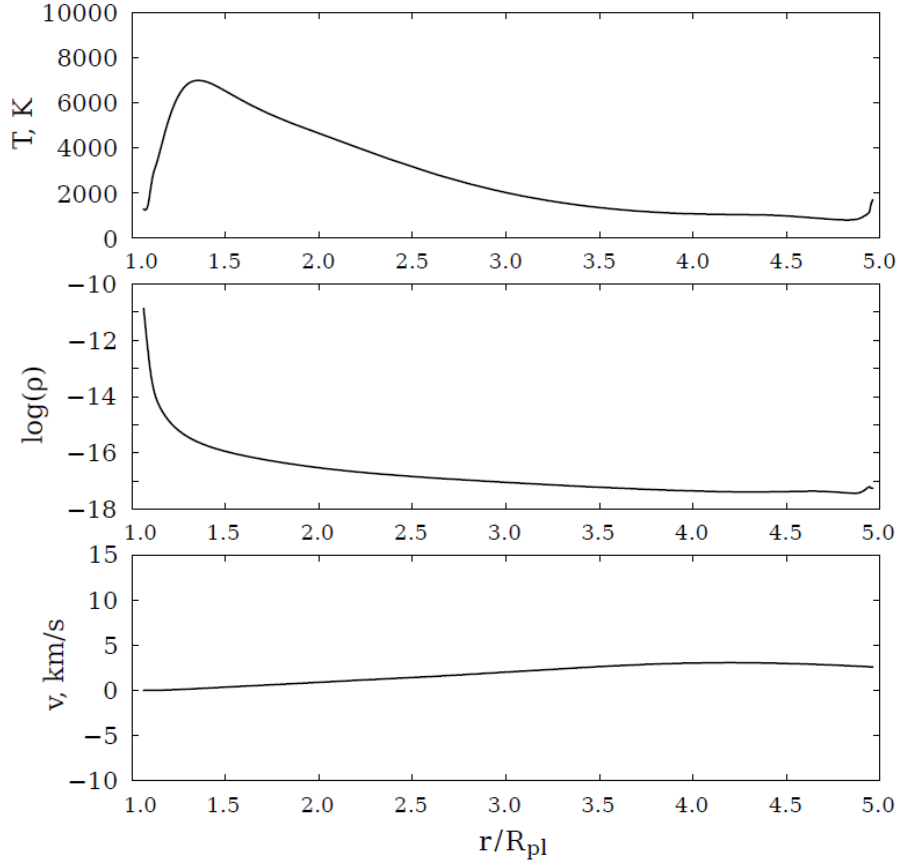

Figure 1. Radial distributions of the temperature (upper panel), mass density in g/cm$^3$ (middle panel) and velocity (lower panel) in the atmosphere of a hot-Jupiter along the "planet-star" line , computed using the aeronomic model [9].

This aeronomic model made it possible to study the dynamic state of HD 209458b's atmosphere and to estimate the effect of reactions with suprathermal photoelectrons on the atmospheric dynamics, chemical composition, and outflow rate [9]. In the present study, this model is used to calculate the dynamic response of the planetary atmosphere to the impact of the host star's flare. The steady state of the hydrogen-helium atmosphere of HD 209458b, for which we assumed amass of $M_{pl}$ =0.69$M_{Jup}$, a radius of $R_{pl}$ = 1.35$R_{Jup,}$ and an orbital semi-major axis of a = 0.045 a.u., is shown in Fig. 1. In particular, it is seen that the maximum temperature is reached at an altitude of 1.3$R_{pl}$ and it is equal to 7000 K. In this solution, a constant hydrodynamic outflow of gas is present [9].

Computations were made for a stellar flare, the parameters of which corresponded to the data of [4,13,14] and were chosen as follows: (a) the stellar radiation in the soft X-rays and in the extreme ultraviolet increases by a factor of 100; (b) the duration of the flare was assumed to be 24 minutes [13]. We further assumed that the high-energy stellar fluxes remained high for the whole duration of the flare. These parameters vary depending on the class of the stellar flare.

RESULTS OF THE COMPUTATIONS

As expected, the exposure of the planetary atmosphere to the stellar flare causes local heating of the atmosphere. Local heating of the atmospheric gas as a result of absorption of the extreme UV radiation of the stellar flare leads to the dynamic response of the atmosphere, namely, to the formation of two shock waves propagating in the atmosphere of the hot exoplanet. It should be noted that the timescale of local heating by atmospheric radiation is shorter than the timescale of the dynamic response to heating and, therefore, assumed a simplified profile of intensity variation during the flare does not affect the results. The change of the dynamic state due to the action of stellar flare is shown in Fig. 2.

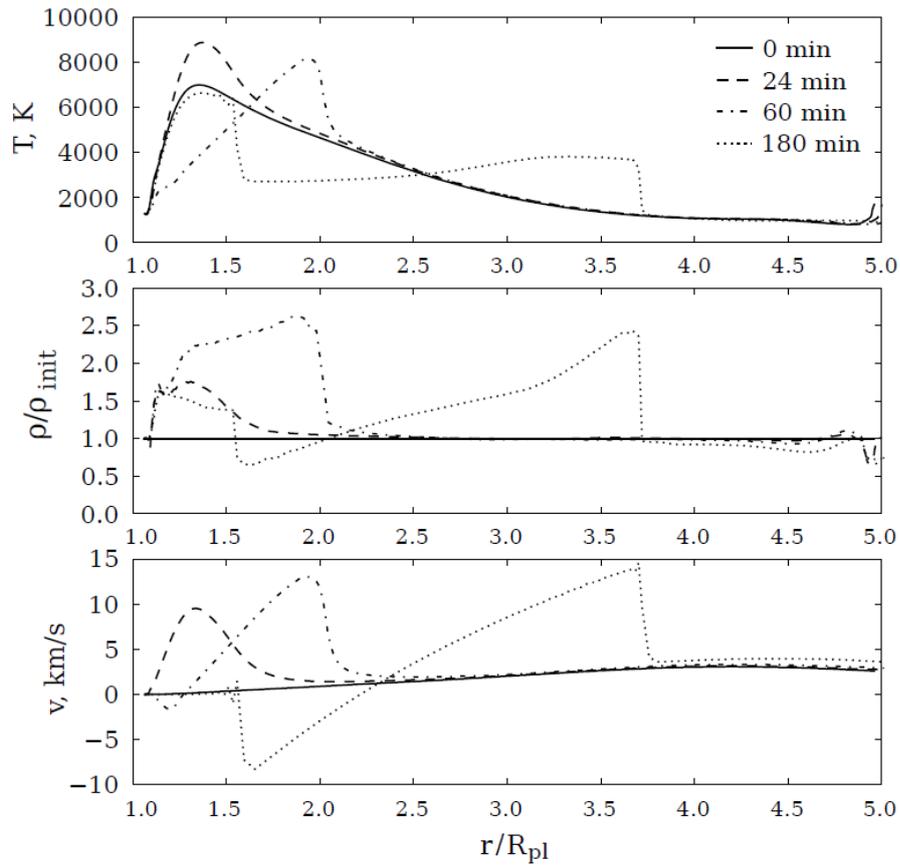

Figure 2. Changes in the dynamic state of the atmosphere due to the action of stellar flare. We show radial profiles of the temperature (top panel), relative mass density (middle panel) and radial velocity (bottom panel) in the atmosphere of the hot Jupiter HD 209458b before the flare (solid line), during the flare (dashed line) and after its pass (at 60 min. - dot-dashed line; at 180 min. - dotted line)

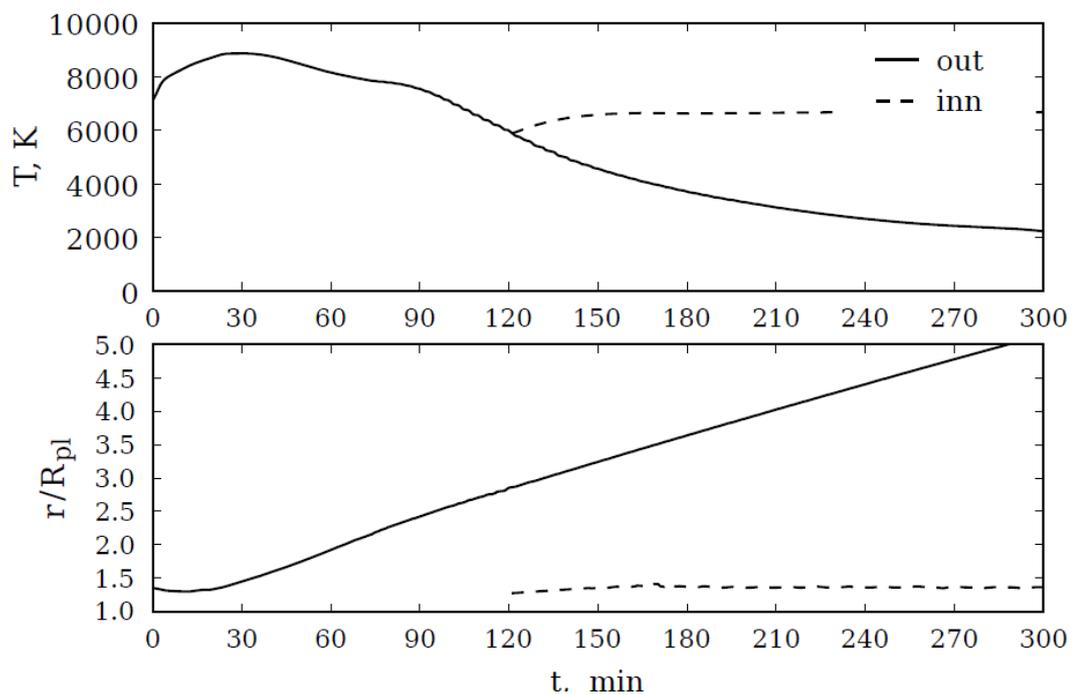

Figure 3. Variations of the peak temperature at radial profile (upper panel) and temperature peak height (bottom panel) since the beginning of the stellar flare. The dashed line shows the parameters after the passage of the internal shock wave, when the solution returns to initial.

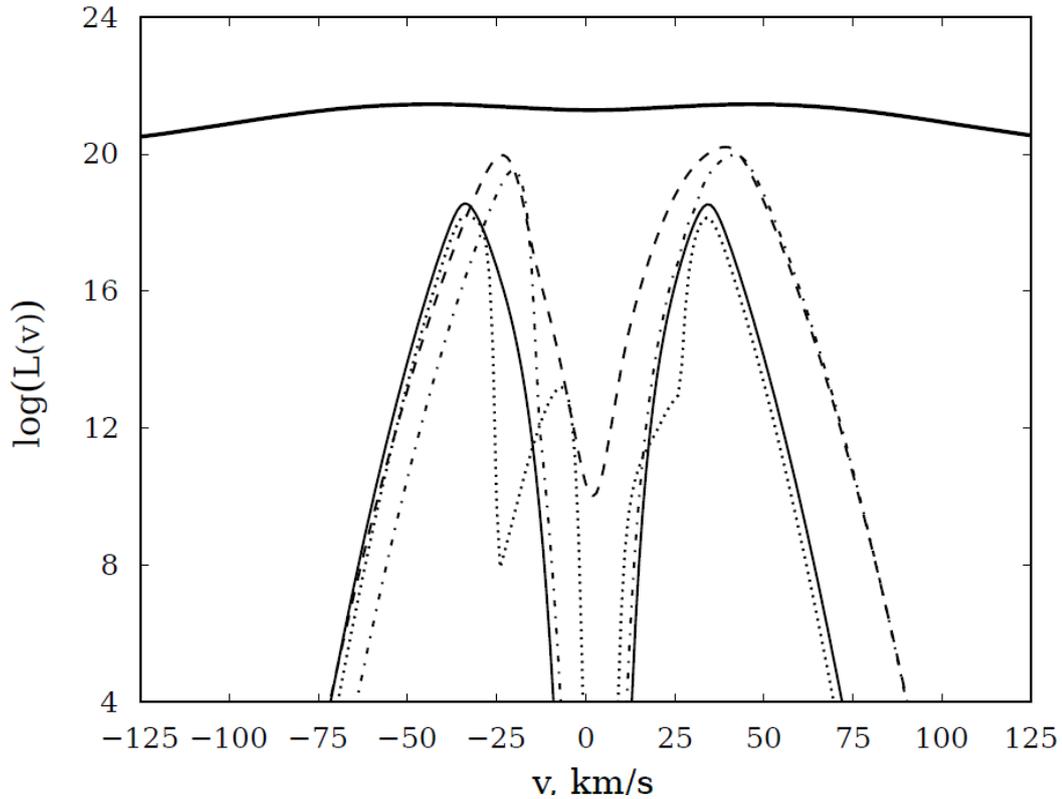

Figure 4. Profiles of the luminosity (in erg/(s×cm/s)) of the atmosphere of the hot-Jupiter HD 209458b in the Ly-α line before the flare (thin continuous line), during the flare (dashed line) and after its completion (at 60 min. - dash-dotted line, at 180 min. - dotted line) . For comparison, we show also the line profile of the host star (thick solid line).

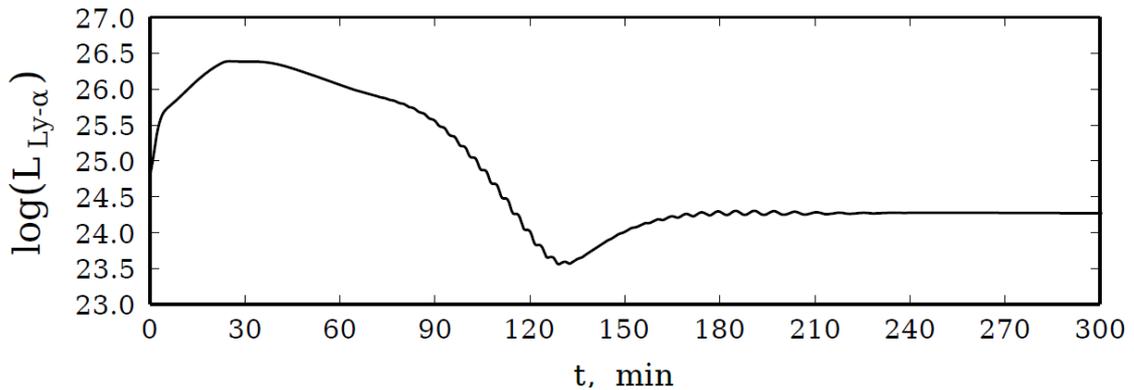

Figure 5. Variations of the total luminosity (in erg/s) of the atmosphere of the hot Jupiter HD 209458b in Ly-α line at its outer boundary before, along, and after the flare event.

As Fig. 2 shows, during the flare the temperature increases without significant shift of the peak temperature position in the atmosphere, reaching at the end of the flare a value of 9000 K. After ~30 min, after completion of the heating, an external shock wave forms and simultaneously the temperature of the gas behind the shock wave decreases due to the adiabatic cooling. Further, an internal shock wave forms. After its passage, the dynamic state of the gas returns to the initial state. The details of the dynamic response of the atmosphere to the stellar flare are shown in Fig. 3, where the changes in the value and height of the peak temperature since the start of the stellar flare are shown. The solid line shows gas parameters behind the outer shock wave and the dashed line shows the parameters of the gas, recovering after the passage of the internal shock wave. As we noted above, during the flare the gas temperature increases by 2000 K locally, while the height of the temperature peak (radial distance) decreases. As the atmosphere starts to react on the impact of a stellar flare , the temperature in the peak decreases, and the temperature minimum is reached in two hours after the onset of the flare. As it can be seen, this time corresponds to the beginning of the restoration of the initial gas state.

In order to assess the possibility of an observational manifestation of the impact of the stellar flare on the upper atmosphere of the planet, we estimated the luminosity of the atmosphere in the Ly-α line. For this, we used results of computations of hydrodynamic characteristics of the atmosphere (Fig. 3), which allowed us to calculate for every cell in the computational domain the source function in the Ly-α line for the local temperature under assumption of local thermodynamical equilibrium (LTE). Further, taking into account the radial distribution of the atomic hydrogen, we estimated the absorption of the radiation in the Ly-α line, which allowed us to calculate the profile of the Ly-α line at the outer boundary of the exoplanet atmosphere. In this case, both for emission and absorption of Ly-α photons, we took into account Doppler shifts caused by the radial gas flow.

Figure 4 shows the luminosity profiles in the Ly-α line before, during and after the impact of the stellar flare. It follows from calculations that at the outer boundary of the planetary atmosphere a double-humped profile of the luminosity of the atmosphere in Ly-α line forms. Figure 5 shows how the total luminosity of the atmosphere in the Ly-α line varies due to the response of the atmosphere to the impact of the stellar flare. The modelled profile is determined by the fact that the radiation of the gas in the region of the temperature peak leads to the formation of a line with extended wings, which are not absorbed by the cooler gas in the overlying layers of the atmosphere. As it can be seen from Fig. 5, immediately after the end of the flare, due to the gas heating, the total luminosity increases approximately by two orders of magnitude. At the same time, the gas begins to move (see Fig. 3, bottom panel), causing the peaks at the Ly-α line profile to shift. At 60 and 180 min. after the flare, the features reflecting the effects of propagation of shock waves in the atmosphere appear in the line profile. Due to the planetary orbital motion, with a velocity of 144 km/s, the profile of the Ly-α line of the planet will periodically shift relative to the profile of the host star line by this value and, accordingly, the luminosity of the planet at such relative shifts becomes comparable to the luminosity of the star. This circumstance can be regarded as one of the possible observational manifestations of the impact of a flare.

We note that, for the steady state of the atmosphere, the total energy flux in the Ly-α line emanating from the atmosphere of the hot-Jupiter HD 209458b is estimated to be about $2 \times 10^{24}$ erg s$^{-1}$. At the same time, the energy flux in this line from the host star is equal to $5 \times 10^{28}$ erg s$^{-1}$. Accordingly, the ratio of the energy fluxes of the planetary atmosphere and of the host star in the Ly-α line is of the order of $10^{-4}$. As it follows from Fig. 5, at the time of the impact this ratio can increase by several orders of magnitude. As modern photon detectors allow to detect fluxes with such a ratio of values, it is possible to observe manifestations of shock waves, characterizing dynamic response of the atmosphere to the action of a stellar flare.

CONCLUSIONS

We presented results of calculations of the dynamic response of the atmosphere of the planet HD 209458b to the impact of the host star flare. It is shown that the absorption of the additional energy of the stellar flare in the extreme ultraviolet leads to a substantial local heating of the atmosphere. One of the consequences of this is the formation of two shock waves, propagating in the atmosphere of the planet. We discussed possible observational manifestations of the shock waves and feasibility of their detection both in the luminosity of the atmosphere of the exoplanet in the Ly-α line profile and in the total energy fluxes in this line from the atmospheres of the exoplanet and of the host star. As well, the impact of the flare should cause an additional loss of matter from the atmosphere due to its dynamic response to the stellar flare.

ACKNOWLEDGMENTS

This study was supported by the Russian Science Foundation (project No. 18-12-0447).

REFERENCES

1. R. Yelle, Icarus **170**, 167 (2004).

2. J.W. Chamberlain, D. Hunten, *Theory of planetary atmospheres. An introduction to their physics and chemistry* (New York: Acad. Press, 1987).

3. D.V. Bisikalo, V.I. Shematovich, Astron. Reports **59**, 836 (2015).

4. J.R.A. Davenport, Astrophys. J. **829**, article id. 23(12 pp.) (2016).

5. R. Luger, R. Barnes, E. Lopez, J. Fortney, B. Jackson, V. Meadows, Astrobiology 15, 57 (2015).

6. M.A. Tilley, A. Segura, V. Meadows, S. Hawley, J. Davenport, arXiv:1711.08484, (2018).

7. D.V. Bisikalo, A.A. Cherenkov, Astron. Reports **60**, 183 (2016).

8. A. Cherenkov, D. Bisikalo, L. Fossati, C. Möstl, Astrophys. J. **846**, 31 (2017).

9. D.E. Ionov, V.I. Shematovich, Ya. N. Pavlyuchenkov, Astron. Reports **61**, 387 (2017).

10. A. Vidal-Madjar, A. Lecavelier des Etangs, J.-M. Desert, G. E. Ballester, R. Ferlet, G. Hebrard, M. Mayor, Nature **422**, 143 (2003).

11. J.R. Kulow, K. France, J. Linsky, R.O.P. Loyd, Astrophys. J. 786, article id. 132(9 pp.) (2014).

12. V.I. Shematovich, D.E. Ionov, H. Lammer, Astron. Astrophys. **57**1, id. A94:7 pp. (2014).

13. A. Veronig, M. Temmer, A. Hanslmeier, W. Otruba, M. Messerotti, Astron. Astrophys. **382**, 1070 (2002).

14. B. T. Tsurutani, D. L. Judge, F. L. Guarnieri, P. Gangopadhyay, A. R. Jones, J. Nuttall, G. A. Zambon, L. Didkovsky, A. J. Mannucci, B. Iijima, R. R. Meier, T. J. Immel, T. N. Woods, S. Prasad, L. Floyd, J. Huba, S. C. Solomon, P. Straus, R. Viereck, Geophys. Res. Lett. **32**, L03S09 (2005).